\documentclass[11pt]{article}

\usepackage{ifthen}
\usepackage{latexsym}
\usepackage{amsfonts}
\usepackage{amsmath}
\usepackage{amssymb}
\usepackage{algpseudocode}

\usepackage{mathtools}
\usepackage{csquotes}
\usepackage{algorithm}

\usepackage{pifont}
\mathtoolsset{centercolon}

\newcommand{\hugeDebug}{false}
\ifthenelse{\equal{\hugeDebug}{false}}{
\newcommand{\normalspacing}{\singlespacing}
\pagestyle{plain}

\setlength{\oddsidemargin}{0.25in}
\setlength{\evensidemargin}{\oddsidemargin}
\setlength{\textwidth}{6in}
\setlength{\textheight}{8in}
\setlength{\topmargin}{-0.0in}
}
{

\newcommand{\normalspacing}{\niceninespacing}
\pagestyle{headings}
 \setlength{\oddsidemargin}{-0.4in}
 \setlength{\evensidemargin}{\oddsidemargin}
 \setlength{\textwidth}{5.3in}
 \setlength{\textheight}{6in}
 \setlength{\textheight}{6.25in}
 \setlength{\topmargin}{-0.0in}
}

\makeatletter
\newcommand{\singlespacing}{\let\CS=
\@currsize\renewcommand{\baselinestretch}{1}\tiny\CS}
\newcommand{\niceninespacing}{\let\CS=\@currsize\renewcommand{\baselinestretch}{1.9}\tiny\CS}
\makeatother

\normalspacing

\newcommand{\pvnp}{P-versus-NP}
\newcommand{\C}{\mathcal{C}}
\makeatletter
\g@addto@macro\bfseries{\boldmath}
\makeatother
\usepackage[raggedright]{titlesec}

\title{A Refutation of Guinea's ``Understanding SAT is in P"}
\author{
Jackson Abascal\thanks{
Work supported in part by a CRA-W CREU grant.
}~~and Shir Maimon\footnotemark[1]
\\Department of Computer Science \\
University of Rochester \\
Rochester, NY 14627, USA
}

\begin{document}

\sloppy

\maketitle

\begin{abstract}
In this work, we summarize and critique the paper ``Understanding SAT is in P" by Alejandro S\'anchez Guinea~\cite{guinea}. The paper claims to present a polynomial-time solution for the NP-complete language 3-SAT. We show that Guinea's algorithm is flawed and does not prove 3-SAT is in P.
\end{abstract}

\section{Introduction}\label{intro}
The problem of finding a satisfying assignment to a boolean formula, referred to as SAT, has strong historical significance in computer science. It was the first natural problem proven to be NP-complete \cite{Cook71thecomplexity} and remains an area of active theoretical and practical interest.

Alejandro S\`anchez Guinea's paper ``Understanding SAT is in P"\footnote{This critique is written with respect to the most recent available revised version: Version 4 (the version of September 16, 2016) of arXiv.org report 1504.00337 \cite{guinea}} claims to provide a polynomial time algorithm for 3-SAT \cite{guinea}, a restricted version of SAT. Like SAT, 3-SAT is NP-complete \cite{Kar72}.

If correct, Guinea's paper would resolve the \pvnp{} problem by showing that P~=~NP. However, we show in Section \ref{refutation} that the algorithm is based on multiple erroneous assumptions and admits counterexamples.

We begin with some definitions. A \textit{literal} is either a boolean variable $x$ or a negated variable $\neg x$.
A \textit{clause} is an expression of literals connected by $\lor$ (logical~or) operators.
We say a clause is \textit{satisfied} if it evaluates to true.
For example, one satisfying assignment of $(x \lor \neg y \lor z)$ is $x=$ True, $y=$ True, and $z=$ False.
The 3-SAT problem is defined as follows: Given a collection of variables and clauses where each clause contains exactly three literals, determine whether there exists an assignment of the variables that satisfies every clause.
Distinct clauses may share variables.

Guinea defines 3-SAT in a somewhat nonstandard fashion by disallowing any clause that contains the same variable twice. This is not an issue, as standard 3-SAT can be reduced in polynomial time to Guinea's form (see Appendix).

Given an instance of 3-SAT, Guinea defines an \textit{understanding} $\tilde{u}$ as a function from the set of literals to the set $\{t,f,\epsilon\}$. For any literal $\lambda$, we say that $\lambda$ is ``true" under $\tilde{u}$ if $\tilde{u}(\lambda) = t$, is ``false" if $\tilde{u}(\lambda) = f$, and is ``free" otherwise. The definition as Guinea presents it does not enforce consistency between literals and their negations. For example, both $x$ and $\neg x$ may be true under $\tilde{u}$.

When evaluating a 3-SAT expression with respect to an understanding $\tilde{u}$, we consider a clause to be satisfied if any of its literals are true under $\tilde{u}$. Note that if an understanding satisfies the clauses of an expression, this does not mean that there exists a satisfying assignment in the traditional sense. We call two understandings \textit{equivalent} if they satisfy the same set of clauses.

Let $\varphi$ be a clause, and for a literal $\lambda$ in $\varphi$ define the \textit{concept} of $\lambda$ in $\varphi$ as the multiset $\{\tilde{u}(l_1),\tilde{u}(l_2)\}$ where $l_1,l_2$ are the two literals in $\varphi$ which are not $\lambda$. A literal $\lambda$ has no concept in $\varphi$ if $\lambda \notin \varphi$.
Although it is never explicitly stated, Guinea assumes that the literals and the clause associated with a concept are implicitly stored and can be recovered.
We say that a concept $\C$ is of type $\C^*$ if it contains the true symbol $t$, and of type $\C^+$ otherwise.
Intuitively, if a concept with respect to $\lambda$ is of type $\C^*$ then $\lambda$ can be set to free, as some other literal already satisfies its associated clause.
If it is of type $\C^+$ then $\lambda$ must evaluate to true in order to satisfy the clause (assuming no other values change).

For any set of clauses $\phi$ and literal $\lambda$, define $\widetilde{\C}[\lambda]$ as the set of concepts $\C$ such that $\C$ is the concept of $\lambda$ in $\varphi$ for some $\varphi \in \phi$. Define $\widetilde{\C}[\lambda]^{-}$ as the set of concepts of type $\C^+$ in $\widetilde{\C}[\neg\lambda]$.
Intuitively, $\widetilde{\C}[\lambda]^-$ is the set of concepts associated to clauses which would be unsatisfied if $\lambda$ were true.
A set of concepts is of type $\widetilde{\C}^*$ if all of the concepts in it are of type $\C^*$, and a set of concepts is of type $\widetilde{\C}^+$ if at least one of the concepts within it are of type $\C^+$.

An understanding is called \textit{defined} if it satisfies the following property:

\[
\tilde{u}(\lambda)= 
\begin{cases*}
\epsilon & if $\widetilde{\C}[\lambda]$ is empty or $(\widetilde{\C}[\lambda]^-\text{ is empty and } \widetilde{\C}[\lambda] \text{ is of type } \widetilde{\C}^*)$ \\
t & if $\widetilde{\C}[\lambda]$ is of type $\widetilde{\C}^+$ and $\widetilde{\C}[\lambda]^-$ is empty \\
f & if $\widetilde{\C}[\lambda]^-$ is not empty and  $\widetilde{\C}[\lambda]$ is not of type $\widetilde{\C}^+$\\
\end{cases*} \]
The case where $\widetilde{\C}[\lambda]$ is of type $\widetilde{\C}^+$ and $\widetilde{\C}[\lambda]^-$ is not empty is purposefully omitted. In this case $\tilde{u}$ is considered undefined.
If an understanding is defined with respect to some set of clauses then $\lambda$ and $\neg \lambda$ will not have conflicting assignments.
In particular, $\tilde{u}(\lambda) = t$ implies $\tilde{u}(\neg \lambda) = f$, and $\tilde{u}(\lambda) = \epsilon$ implies $\tilde{u}(\neg \lambda) = \epsilon$.
A set of clauses $\phi$ is satisfiable if and only if there is some understanding defined with respect to $\phi$. 
Guinea's algorithm claims to produce a defined understanding for the clauses of an instance of 3-SAT if and only if the instance is satisfiable.

Observe that although the existence of a satisfying assignment implies the existence of a defined understanding, this understanding will not necessarily assign to each literal the value given by the satisfying assignment.
In a defined understanding, a literal $\lambda$ maps to $t$ only if there is a clause in which $\lambda$ is the only literal assigned to true.
Furthermore, $\lambda$ can only be assigned to $f$ if there is a clause in which $\neg \lambda$ would be the only literal assigned to true. In any other case, $\lambda$ must be free.
For example, with respect to the 3-SAT instance consisting of the single clause $(x \lor y \lor z)$, the understanding that assigns each of $x$, $y$, and $z$ to $t$ is not defined, whereas the understanding that assigns $x$ to $t$ and both $y$ and $z$ to $\epsilon$ is considered defined.

\section{Algorithms}
Guinea's descriptions of his algorithms are often difficult to read and occasionally ambiguous. We have written in this section a more precise exposition of each of the algorithms presented in the paper.

Guinea presents three algorithms: Algorithm G, Algorithm D, and Algorithm~${\rm \widetilde{U}}$\@. Algorithm ${\rm \widetilde{U}}$ is claimed to ultimately solve 3-SAT in polynomial time. He also defines an operation $\langle$Compute $\tilde{u} \rangle$, which is used in all three algorithms. We will describe $\langle$Compute~$\tilde{u}\rangle$ first.

\subsection{$\langle$Compute $\tilde{u}\rangle$}
The exact description of $\langle$Compute $\tilde{u}\rangle$ in the paper is as follows:
\begin{displayquote}
``$\langle$Compute $\tilde{u}\rangle$ =
    Compute $\tilde{u}$ for each literal $\lambda$ and its negation for which the type $\widetilde{\C}[\lambda ]$ has changed, until there is no change of type on any subset of concepts of $\widetilde{\C}$" \cite{guinea}.
\end{displayquote}

We found this description ambiguous. Because Guinea calls $\langle$Compute $\tilde{u}\rangle$ an operation rather than an algorithm, we believe he means that we can insert its description anywhere he writes ``$\langle$Compute $\tilde{u}\rangle$" and interpret the description in a way that makes sense in context.
This would imply that for each $\langle$Compute $\tilde{u}\rangle$ call, the first literals acted on are those for which the type of $\widetilde{\C}[\lambda]$ has changed in the operations directly preceding the call. Thus we can consider $\langle$Compute~$\tilde{u}\rangle$ as a function taking in some set of literals whose values under $\tilde{u}$ we would like to modify.

Furthermore, it is unclear what it means to ``Compute $\tilde{u}$ for a literal $\lambda$."
We have taken this phrase to be equivalent to the procedure $\Call{Recalculate}{\tilde{u},\widetilde{\C},\lambda}$ defined below, where $\tilde{u}$ is an understanding, $\widetilde{\C}$ is a set of concepts, and $\lambda$ is a literal. 

\begin{algorithmic}[1]
  \Procedure{Recalculate}{$\tilde{u}, \widetilde{\C}, \lambda$}
  \If{$\widetilde{\C}[\lambda]$ is empty \textbf{or} $(\widetilde{\C}[\lambda]^-\text{ is empty \textbf{and} } \widetilde{\C}[\lambda] \text{ is of type } \widetilde{\C}^*)$}
    \State \Return $\epsilon$
  \ElsIf{$\widetilde{\C}[\lambda]$ is of type $\widetilde{\C}^+$ \textbf{and} $\widetilde{\C}[\lambda]^-$ is empty}
    \State \Return $t$
  \ElsIf{$\widetilde{\C}[\lambda]^-$ is not empty \textbf{and}  $\widetilde{\C}[\lambda]$ is not of type $\widetilde{\C}^+$}
    \State \Return $f$
  \EndIf
  \State \Return UNDEFINED
  \EndProcedure
\end{algorithmic}

We can now define our interpretation of $\langle$Compute $\tilde{u}\rangle$. We have defined it as a function that takes as arguments an understanding $\tilde{u}$, a set of concepts $\widetilde{\C}$, and an initial set $S$ of literals to be recomputed.

\begin{algorithmic}[1]
  \Procedure{Compute}{$\tilde{u},\widetilde{\C},S$}
    \While{$S$ is not empty}
       \State Take any $\lambda\in S$ and remove $\lambda$ from $S$
       \State Set $\tilde{u}(\lambda):=$ \Call{Recalculate}{$\tilde{u}, \widetilde{\C}, \lambda$}
       \If{$\tilde{u}(\lambda)=$ UNDEFINED}
        \State \Return UNDEFINED
       \EndIf
      \State For any concept $\C$ in $\widetilde{\C}$ containing $\lambda$, update $\C$ using $\tilde{u}$\label{algc1}
       \ForAll{literals $\lambda'$ for which the type of $\widetilde{\C}[\lambda']$ changed in line \ref{algc1}}
           \State Add $\lambda'$ to $S$ \Comment $\widetilde{\C}$ changed, so there may be literals whose concepts changed
       \EndFor
    \EndWhile
    \State \Return ($\tilde{u},\widetilde{\C}$)
  \EndProcedure
\end{algorithmic}
$\langle$Compute $\tilde{u}\rangle$, while ambiguously defined, is not the basis of our refutation. Under any reasonable interpretation of $\langle$Compute $\tilde{u}\rangle$, our counterexamples remain valid.

\subsection{Algorithm G}
Algorithm G takes as arguments a set of clauses $\phi$, an understanding $\tilde{u}$, a set of concepts $\widetilde{\C}$ defined with respect to $\tilde{u}$, and a literal $\lambda$ that is free under $\tilde{u}$.
Let $\phi_{\lambda} \subseteq \phi$ be the clauses in which $\lambda$ or $\neg \lambda$ appears. 
Guinea claims the algorithm will return ``True" if and only if there is an understanding $\tilde{u}_\lambda$ defined with respect to $\phi_\lambda$ such that $\tilde{u}_\lambda(\lambda) = t$.
Guinea does not specify how exactly to assign $\tilde{u}(l_1)$ and $\tilde{u}(l_2)$ in line \ref{algg1}, meaning the algorithm is ill-defined.
Due to Algorithm G's heavy reliance on $\langle$Compute $\tilde{u}\rangle$, rather than analyzing this algorithm our refutation will assume its validity.

\begin{algorithmic}[1]
\Procedure{Algorithm\_G}{$\phi,\tilde{u},\widetilde{\C},\lambda$}
    \State Set $\tilde{u}(\lambda)\coloneqq t$
    \State Set $\tilde{u}(\neg \lambda) \coloneqq f$
    \ForAll{concepts $\C$ in $\widetilde{\C}[\lambda]$}
        \State Let $l_1$ and $l_2$ be the literals in $\C$
        \State Set both $\tilde{u}(l_1)$ and $\tilde{u}(l_2)$ to either $\epsilon$ or $f$ \label{algg1}
        \State Let $S$ be the literals $\lambda$ such that $\widetilde{\C}[\lambda]$ changed type in the previous step
        \State Set $(\tilde{u}',\widetilde{\C}') \coloneqq$ \Call{Compute}{$\tilde{u},\widetilde{\C},S$}
        \If {$(\tilde{u}',\widetilde{\C}') \neq $ UNDEFINED}
            \State \textbf{return} True
            \Comment The algorithm terminates successfully 
        \EndIf
    \EndFor
    \State \textbf{return} False \Comment The algorithm terminates unsuccessfully
\EndProcedure
\end{algorithmic}

\subsection{Algorithm D}
Algorithm D claims to do the following: Given an understanding $\tilde{u}$, a set of concepts $\widetilde{C}$, and a literal $\lambda$ that is false under $\tilde{u}$, construct a defined understanding $\tilde{u}'$ equivalent to $\tilde{u}$ such that $\lambda$ is free under $\tilde{u}'$, or report that no such $\tilde{u}'$ exists.
Guinea refers to a set $\mathcal{H}$ without explicitly specifying its purpose, saying only $\mathcal H$ is ``considered empty, if not given" \cite{guinea}.
To properly call Algorithm D recursively, we modify the definition to be dependent on $\mathcal{H}$ by adding the condition that $\tilde{u}'(\lambda) = \tilde{u}(\lambda)$ if $\lambda \in \mathcal{H}$.

\begin{algorithmic}[1]
  \Procedure{Algorithm\_D}{$\tilde{u},\widetilde{\C},\lambda,\mathcal{H}$}
    \While{there is a concept in $\widetilde{\C}[\lambda]^-$ not already considered} \label{algd1}
      \State Take any $\C \in \widetilde{\C}[\lambda]^-$ not already considered
      \State Let $Q$ be the set of literals $l$ in $\C$ such that $l \notin\mathcal H$
      \While{$Q$ is not empty} \label{algd2} 
        \State Take any $l \in Q$ and remove $l$ from $Q$ \label{algd5}
        \State Set $(\tilde{u}',\widetilde{\C}') = $ FAIL
        \If{$\tilde{u}(l) = f$}
             \State Set $(\tilde{u}',\widetilde{\C}') \leftarrow $\Call{Algorithm\_D}{$\tilde{u},l,\mathcal{H}\cup\{\lambda\}$} \label{algd6}
              \If{$(\tilde{u}',\widetilde{\C}') =$ FAIL}
                \State \textbf{go to}~line~\ref{algd2}
              \EndIf
                
        \EndIf
        \If{$\Call{Algorithm\_G}{\tilde{u},l}$}
            \If{$(\tilde{u}',\widetilde{\C}') \neq $ FAIL}
                \State Set $\tilde{u} \coloneqq \tilde{u}'$
                \State Set $\widetilde{\C} \coloneqq \widetilde{\C}'$
            \EndIf
            \State Set $\tilde{u}(l) \coloneqq t$ \label{algd3}
            \State Set $\tilde{u}(\neg l) \coloneqq f$ \label{algd4}
            \State Let $S$ be the literals $l'$ such that $\widetilde{\C}[l']$ changed type in lines \ref{algd3} and \ref{algd4}
            \State ($\tilde{u},\widetilde{\C}$) $\coloneqq$ \Call {Compute} {$\tilde{u},\widetilde{\C},S$} \label{algd8}
            \State \textbf{go to}~line~\ref{algd1}
        \EndIf
      \EndWhile
      \State \Return FAIL
    \EndWhile
    \State \Return $(\tilde{u},\widetilde{\C})$ \label{algd7}
  \EndProcedure
\end{algorithmic}

\subsection{Algorithm ${\rm \widetilde{U}}$}
Algorithm ${\rm \widetilde{U}}$ takes a 3-SAT problem instance $\Phi$ as its only argument. Guinea claims the algorithm will produce an understanding $\tilde{u}$ defined with respect to the clauses of $\Phi$, or report that none exist.

\begin{algorithmic}[1]
  \Procedure{Algorithm\_${\rm \widetilde{U}}$}{$\Phi$}
    \State Let $\tilde{u}$ be an understanding assigning every literal in $\Phi$ to $\epsilon$
    \State Let $\phi$ be $\varnothing$
    \State Let $\widetilde{\C}$ be $\varnothing$
    \While{there is some clause in $\Phi$ not in $\phi$}
      \State Let $\varphi$ be a clause in $\Phi$ but not $\phi$ \label{algu2}
      \If{all literals $\lambda \in \varphi$ are false under $\tilde{u}$}\label{algu4}
        \ForAll{literals $\lambda \in \varphi$} \label{algu3}
            \State Set $(\tilde{u}',\widetilde{\C}')\coloneqq$ \Call{Algorithm\_D}{$\tilde{u},\widetilde{\C},\lambda,\varnothing$}\label{algu5}
            \If{$(\tilde{u}',\widetilde{\C}')\neq$ FAIL}
                \State Set $\tilde{u}\coloneqq\tilde{u}'$
                \State Set $\widetilde{\C}\coloneqq\widetilde{\C}'$
                \State \textbf{go to}~line~\ref{algu1}
            \EndIf
        \EndFor
        \State \textbf{return} FAIL \Comment{$\Phi$ is unsatisfiable}\label{algu6}
      \EndIf
      \ForAll{literals $\lambda$ in $\varphi$ (first taking literals that are not false under $\tilde{u}$)} \label{algu1}
        \State Add the concept of $\lambda$ in $\varphi$ to $\widetilde{\C}$
        \State Let $S$ be the set of literals $l$ such that $\widetilde{\C}[l]$ changed type in the previous step
        \State ($\tilde{u},\widetilde{\C}$) $\coloneqq$  \Call{Compute}{$\tilde{u},\widetilde{\C},S$}
        \State Add $\varphi$ to $\phi$
      \EndFor
    \EndWhile
    \State \textbf{return} $\tilde{u}$ \Comment{We have found a satisfying assignment for $\Phi$}
  \EndProcedure
\end{algorithmic}

\section{Refutation}\label{refutation}

\subsection{Refutation of Algorithm D}

A major error in Algorithm D stems from the fact that concepts removed from $\widetilde{\C}[\lambda]^-$ throughout the execution of the algorithm can be readded. Because Guinea states that we only consider each concept at most once, some concept that is first removed from $\widetilde{\C}[\lambda]^-$ and then later added to $\widetilde{\C}[\lambda]^-$ will not be processed a second time. We can use this fact to construct a case in which the algorithm terminates successfully, even though there is no defined understanding in which $\lambda$ is free.

Consider the following instance of 3-SAT:
\begin{align*}
(a \lor b \lor c) \land (a \lor \neg b \lor c) \land (a \lor b \lor \neg c) \land (a \lor \neg b \lor \neg c)\land 
\\(\neg x \lor y \lor \neg a) \land (\neg x \lor \neg y \lor \neg a) \land (d \lor y \lor e) \land (\neg d \lor \neg y \lor \neg e).
\end{align*}
and some initial defined understanding $\tilde{u}$ which satisfies $\tilde{u}(x) = f$, $\tilde{u}(a) = t$, and $\tilde{u}(y) = t$.
An example is the understanding $\tilde{u}$ defined by the following assignments:
\begin{align*}
\begin{array}{cccc}
    \tilde{u}(a)=t, &\tilde{u}(\neg a)=f,
    &\tilde{u}(b)=\epsilon, &\tilde{u}(\neg b)=\epsilon, \\
    \tilde{u}(c)=\epsilon, &\tilde{u}(\neg c)=\epsilon,
    &\tilde{u}(d)=f, &\tilde{u}(\neg d)=t, \\
    \tilde{u}(e)=\epsilon, &\tilde{u}(\neg e)=\epsilon,
    &\tilde{u}(x)=f, &\tilde{u}(\neg x)=t, \\
    \tilde{u}(y)=t, &\tilde{u}(\neg y)=f. & &
\end{array}
\end{align*}

Calling $\textproc{Algorithm\_D}(\tilde{u},\widetilde{\C},x,\varnothing)$ should return an understanding under which $x$ is free or report that one does not exist.
The literal $a$ is forced true by the first four clauses, and therefore either $y \lor \neg a$ or $\neg y \lor \neg a$ is false. By the fifth and sixth clauses $\neg x$ must be true and $x$ must be false. Thus no understanding defined with respect to these clauses can assign $x$ to $\epsilon$, and the algorithm should terminate unsuccessfully.

We will simulate a run of $\textproc{Algorithm\_D}(\tilde{u},\widetilde{\C},x,\varnothing)$. On line \ref{algd1}, the algorithm takes a concept in $\widetilde{\C}[x]^-$. The only such concept is the concept of $\neg x$ in $(\neg x \lor \neg y \lor \neg a)$. 
Next, on line \ref{algd5}, we choose any literal in this concept. Suppose we choose $\neg y$.
Assuming Algorithm D is correct, our recursive call
\mbox{\textproc{Algorithm\_D($\tilde{u},\widetilde{\C},\neg y,\{x\}$)}} will find that there \textit{does} exist an understanding $\tilde{u}'$ in which $x$ is not changed and $\neg y$ evaluates to free. An example is the understanding $\tilde{u}'$ defined by the following assignments:
\begin{align*}
\begin{array}{cccc}
    \tilde{u}'(a)=t, &\tilde{u}'(\neg a)=f,
    &\tilde{u}'(b)=\epsilon, &\tilde{u}'(\neg b)=\epsilon, \\
    \tilde{u}'(c)=\epsilon, &\tilde{u}'(\neg c)=\epsilon,
    &\tilde{u}'(d)=f, &\tilde{u}'(\neg d)=t, \\
    \tilde{u}'(e)=t, &\tilde{u}'(\neg e)=f,
    &\tilde{u}'(x)=f, &\tilde{u}'(\neg x)=t, \\
    \tilde{u}'(y)=\epsilon, &\tilde{u}'(\neg y)=\epsilon. & &
\end{array}
\end{align*}
Assuming Algorithm G is correct, it will likewise report that there exists an understanding $\tilde{u}'$ with respect to the clauses containing $y$ or $\neg y$ in which $\neg y$ is true, for example the understanding $\tilde{u}'$ defined by the following assignments:
\begin{align*}
\begin{array}{cccc}
    \tilde{u}'(a)=t, &\tilde{u}'(\neg a)=f,
    &\tilde{u}'(b)=\epsilon, &\tilde{u}'(\neg b)=\epsilon, \\
    \tilde{u}'(c)=\epsilon, &\tilde{u}'(\neg c)=\epsilon,
    &\tilde{u}'(d)=t, &\tilde{u}'(\neg d)=f, \\
    \tilde{u}'(e)=\epsilon, &\tilde{u}'(\neg e)=\epsilon,
    &\tilde{u}'(x)=f, &\tilde{u}'(\neg x)=t, \\
    \tilde{u}'(y)=f, &\tilde{u}'(\neg y)=t. & &\\
\end{array}
\end{align*}
At this point, the Algorithm D will set $\tilde{u}$ to the understanding returned from the recursive call to itself on line \ref{algd6}, and then set $\tilde{u}(y) = f$ on line \ref{algd4}.
The \textproc{Compute} call on line \ref{algd8} will not have any effect, since $\tilde{u}$ will be defined.
When we began, we required only that $\tilde{u}(x) = f$, $\tilde{u}(a) = t$, and $\tilde{u}(y) = t$.
At this point $\tilde{u}(x)$ is still equal to $f$ since $\tilde{u}(x)$ was guaranteed not to change by the recursive call. Similarly, $\tilde{u}(a)$ is still equal to $t$ since it must be true in any defined understanding. However, $\tilde{u}(\neg y)$ is now equal to $t$.
Note that the 3-SAT formula given stays the same if we swap $\neg y$ and $y$. Now, $(\neg x \lor y \lor \neg a)$ is the only clause in $\widetilde{\C}[x]^{-}$.
By the exact argument given above, after running one more iteration of the while loop starting on line \ref{algd1}, we will reach a state satisfying our initial conditions that $\tilde{u}(x) = f$, $\tilde{u}(a) = t$, and $\tilde{u}(y) = t$.
The only clause in $C[x]^-$ has already been considered, and so the algorithm halts and returns successfully on line \ref{algd7}.

We have already shown that there is no defined understanding in which $x$ is free, so Algorithm D returns an incorrect result.
If we try to fix this issue by allowing the same clause to be processed multiple times, this example also shows that Algorithm D can return to a previous state and run forever.

\subsection{Refutation of Algorithm ${\rm \widetilde{U}}$}
Even if Algorithm D were correct, Algorithm ${\rm \widetilde{U}}$ could still fail.

Algorithm $\rm \widetilde{U}$ maintains a set of satisfied clauses $\phi$. The algorithm will choose some clause $\varphi$ not in $\phi$ and attempt to satisfy $\varphi$ while keeping all clauses in $\phi$ satisfied. The algorithm will then add $\varphi$ to $\phi$, and continue this process until $\phi$ contains all clauses in $\Phi$ or we fail in satisfying some clause.

The algorithm may fail when $\varphi$ has every literal assigned false. Guinea claims that if there is a defined understanding that satisfies all clauses in $\Phi$, then

\begin{displayquote} ``The call to Algorithm D $\dots$ defines
successfully an understanding with respect to $\phi$ under which one literal $\lambda$ in $\phi$ is free" \cite{guinea}.
\end{displayquote}

We believe the second $\phi$ in this quote is intended to be $\varphi$, as this is what Algorithm D claims to return in this context. However, Guinea's assumption is incorrect.

Consider the following instance of 3-SAT :
\begin{align*}
&(x\lor a \lor b)\land(x\lor\neg a \lor b)\land(\neg x\lor a \lor\neg b)\land(\neg x\lor\neg a \lor\neg b)\land\\
&(y\lor c \lor d)\land(y\lor\neg c \lor d)\land(\neg y\lor c \lor\neg d)\land(\neg y\lor\neg c \lor\neg d)\land\\
&(z\lor e \lor f)\land(z\lor\neg e \lor f)\land(\neg z\lor e \lor\neg f)\land(\neg z\lor\neg e \lor\neg f)\land\\
&(x \lor y \lor z).
\end{align*}

Depending on the order that the algorithm chooses clauses in line \ref{algu2} and literals in line \ref{algu3}, it is possible that the program reaches a state in which
\begin{alignat*}{2}
    \phi&=\{ \parbox[t]{10cm}{\ensuremath{(x\lor a \lor b),(x\lor\neg a \lor b),(\neg x\lor a \lor\neg b),(\neg x\lor\neg a \lor\neg b), \\
       (y\lor c \lor d),(y\lor\neg c \lor d),(\neg y\lor c \lor\neg d),(\neg y\lor\neg c \lor\neg d), \\
    (z\lor e \lor f),(z\lor\neg e \lor f),(\neg z\lor e \lor\neg f),(\neg z\lor\neg e \lor\neg f)
        \},}} \\ 
   \varphi&=(x \lor y \lor z), \\
    \tilde{u}(x)&=f,\\
    \tilde{u}(y)&=f,\\
    \tilde{u}(z)&=f.
\end{alignat*}
In this case, the condition on line \ref{algu4} will hold true because all literals in $\varphi$ are false under $\tilde{u}$.
There is no satisfying assignment for the clauses in $\phi$ in which any of $x$, $y$ or $z$ is free.
To see this let $x$ be free.
Then to satisfy all clauses we need to satisfy $(a \lor b)\land(\neg a \lor b)\land(a\lor\neg b)\land(\neg a \lor\neg b)$, but this is impossible. It is clear that analogous statements hold for $y$ and $z$.
Thus Algorithm D will always return FAIL in line \ref{algu5} of Algorithm ${\rm \widetilde{U}}$ (assuming correctness of Algorithm D), so Algorithm $\rm \widetilde{U}$ will return FAIL in line \ref{algu6}. However, the 3-SAT instance given is satisfiable via the assignment:
\begin{align*}
    x&=\text{True},\\
    a&=\text{True},\\
    b&=\text{False},\\
    y&=\text{True},\\
    c&=\text{True},\\
    d&=\text{False},\\
    z&=\text{True},\\
    e&=\text{True},\\
    f&=\text{False}.
\end{align*}
Therefore Algorithm $\rm \widetilde{U}$ does not solve 3-SAT as claimed.

\section{Conclusion}
We have given counterexamples for both Algorithm D and Algorithm ${\rm \widetilde{U}}$, even assuming all methods they call work as described by Guinea.
These two algorithms are fundamental flaws in Guinea's approach to solving 3-SAT in polynomial time. Guinea has not shown that 3-SAT is in P and has not resolved P-versus-NP.

\paragraph*{Acknowledgments} We thank Professor Lane A. Hemaspaandra, Michaela Houk, and Jesse Stern for their comments on previous drafts of this paper.

\bibliographystyle{alpha}
\bibliography{gry-creu.bib}

\appendix
\section{Appendix. Reduction from 3-SAT to 3-SAT with No Duplicate Variables in Any Clause} \label{appendix}

Given an instance of $\Phi$ 3-SAT, we would like to reduce it in polynomial time to some other instance $\Phi$' that is satisifiable if and only if $\Phi$ is satisfiable, where no clause in $\Phi$' contains the same variable twice.

Take any clause $\varphi$ in $\Phi$ that contains a duplicate variable $x$.
If both $x$ and $\neg x$ are in $\varphi$, then we remove $\varphi$ from the collection of clauses completely as it is trivially always satisfied.
Otherwise, we may replace a duplicate $x$ or $\neg x$ with some variable that always evaluates to false, and this will not affect satisfiability under any assignment.
Take any symbols $a$, $b$, $c$ not already appearing in any clause and add the clauses $(a \lor b \lor c)$, $(a \lor b \lor \neg c)$, $(a \lor \neg b \lor c)$, and $(a \lor \neg b \lor \neg c)$. Then replace the duplicate literal with $\neg a$.
No matter what the values of $b$ and $c$ are, in one of our four new clauses $a$ will be surrounded by two literals evaluating to false, so $a$ must always evaluate to true.
Thus by replacing the duplicate literal with $\neg a$, we are not affecting the satisfiability of our formula but we are strictly reducing the number of duplicate variables.
Repeating this process until no duplicate variables remain yields our formula $\Phi$'.
\end{document}